# Magic Numbers for Packing Adamantane in Helium Droplets: Cluster Cations, Dications and Trications

M. Goulart[1], M. Kuhn[1], L. Kranabetter[1], A. Kaiser[1], J. Postler[1], M. Rastogi[1], A. Aleem[2], B. Rasul[3], D.K. Bohme[4],* and P. Scheier[1],*

[1] Institut für Ionenphysik und Angewandte Physik, Universität Innsbruck, Technikerstr. 25, A-6020 Innsbruck, Austria

[2] LINAC Project, PINSTECH, P.O. Box Nilore, Islamabad 44000, Pakistan

[3] University of Sargodha, 40100 Sargodha, Pakistan

[4] Department of Chemistry, York University, Toronto, ON M3J 1P3, Canada

* dkbohme@yorku.ca, paul.scheier@uibk.ac.at

**Abstract:** We report the first observation of cations, dications and trications of large clusters of adamantane. Cluster formation was initiated near 0 K in helium droplets and ionization was achieved with one or more collisions with energetic He species (He*, He$^+$ or He*$^-$). The occurrence of Coulomb explosion appeared to discriminate against the formation of small multiply charged clusters. High resolution mass spectrometry revealed the presence of "magic number" m/z peaks that can be attributed to the packing of adamantane molecules into cluster structures of special stability involving preferred arrangements of these molecules. These abundance anomalies were seen to be independent of charge state. While some dehydrogenation of adamantane and its clusters was seen as well, no major transformations into adamantoids or microdiamonds were observed.

# 1. INTRODUCTION

Adamantane, with its highly symmetric cage structure (Figure 1) and unique thermal stability, has become a popular hydrocarbon molecule in a wide range of applications in material and polymer science, molecular electronics, biomedical sciences and chemical synthesis.[1-10] Furthermore, adamantane has achieved a reputation as the lowest member of the diamondoid family of molecules that are found naturally in such extreme and diverse environments as petroleum at depths accessed by commercial oil wells, where their thermal stability can serve to assess the extent of natural oil cracking,[11,12] and apparently in dense interstellar clouds that contain microdiamonds where adamantane molecules may play a role as building blocks.[13-18] Adamantane is the simplest polycyclic saturated hydrocarbon with a cage-like skeleton characteristic of the diamond lattice and so also is intrinsically interesting.

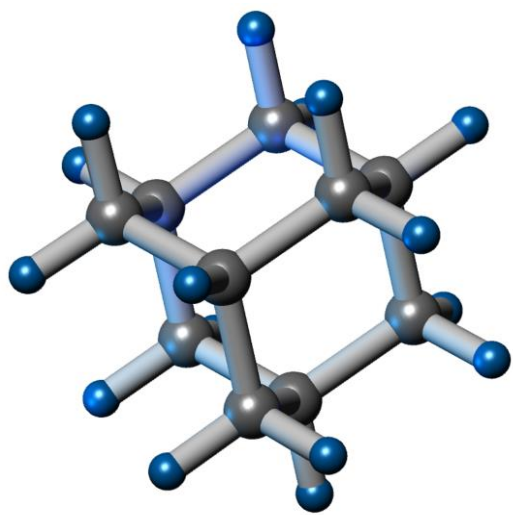

Figure 1. Ball and stick model of the adamantane molecule $C_{10}H_{16}$ with four CH and six $CH_2$ groups.

Here we explore cationic cluster formation with adamantane with the intent to provide insight into the ability of these completely aliphatic rigid cages to pack themselves into preferred arrangements of enhanced stability, or "magic number" clusters, including perhaps arrangements that resemble diamondoids. The observation of cationic clusters of adamantane has not been reported in the literature previously. In our experiments we introduce adamantane molecules into He nanodroplets where they aggregate near 0 K and then expose the aggregates within the droplets to ionization by energetic helium ions or metastables (produced by impacting the He droplets with energetic electrons). The cations that are formed escape the droplets as the He atoms evaporate and then are identified with high-resolution mass spectrometry. The approach and methodology is similar to that adopted in our recent study of the packing in helium nanodroplets of methane, the smallest hydrocarbon molecule.[19] The mass spectra with methane provided clear evidence for magic number cations $(CH_4)_n^+$ with n = 14, 21, 23, 26, 29, 54 and 147 but these were difficult to relate to specific cluster structures, albeit the magic number at n = 54 was consistent with an icosahedral structure containing a vacancy at its center. Furthermore, formation of the dication clusters $(CH_4)_n^{2+}$ was observed for the first time with a threshold at n = 70.

An overview of available results for magic cluster formation within and outside helium droplets will be provided in an upcoming review article (Mauracher, A.; Echt, O.; Ellis, A.M.; Yang, S.; Bohme, D.K.; Postler, J.; Kaiser, A.; Denifl, S.; Scheier, P. Cold Collisions Inside He Nanodroplets: Reactions Close to Zero K. Phys. Rep. 2017). Included are clusters of the noble gases Ar and Kr, the alkali metal atoms Na and K, the metal atoms Mg, Cd, Zn, Ag, In and Cr, and the molecules $H_2O$ and serine, as well as $H_2$ in $(H_2)_nH_3^+$, $CH_4$ in $(CH_4)_nCH_5^+$, and NaF in $(NaF)_nNa^+$. Interesting to note is that the small mixed clusters of adamantane cations, $C_{10}H_{16}^+ \cdot He_{1,2,3}$ and $C_{10}H_{16}^+ \cdot N_2$, have been generated and observed previously by others in studies of the infrared spectrum of the adamantane cation.[20] These cluster ions were generated in a pulsed supersonic plasma beam expansion of adamantane by electron ionization followed by three-body aggregation of either He or $N_2$ with a rotational temperature of less than 50 K.

**EXPERIMENTAL DETAILS**

Helium droplets were produced as a continuous beam by expanding ultra-pure 4He (99.9999%) at a stagnation pressure of 2.5 MPa through a 5 μm diameter aperture into a vacuum. The nozzle is mounted on a copper cylinder which is cooled to 9.5-10 K by a closed-cycle cryostat (Sumitomo RDK-415 F50H). Expansion conditions in the current work were chosen to yield helium nanodroplets with a mean size between $10^5$ and a few droplets with $10^6$ He atoms.[21,22] After passing a conical skimmer (diameter 0.8 mm) to avoid shock fronts, the beam of He nanodroplets enters a differentially pumped chamber and traverses a heated pick-up cell (405 K) containing the vapor of adamantane. The sample was introduced via a heated tube from a reservoir attached to the vacuum chamber and kept at a temperature of 313 K. After the pick-up of adamantane molecules, the droplets passed through a second differentially pumped pickup chamber, allowing the option of adding a second dopant. Finally, the droplets passed through a second skimmer and entered another differentially pumped chamber, where they were exposed to an electron beam of variable energy (0-150 eV). Any positively charged ions produced were then extracted into a high resolution and high repetition rate reflectron time-of-flight mass spectrometer (Tofwerk). Detailed information can be found in refs.[23,24]

## RESULTS AND DISCUSSION

**Ionization of the Adamantane Molecule.** Mass spectra were obtained first for pure adamantane vapor, with and without embedding the adamantane into He nanodroplets. These spectra are shown in Figure 2.

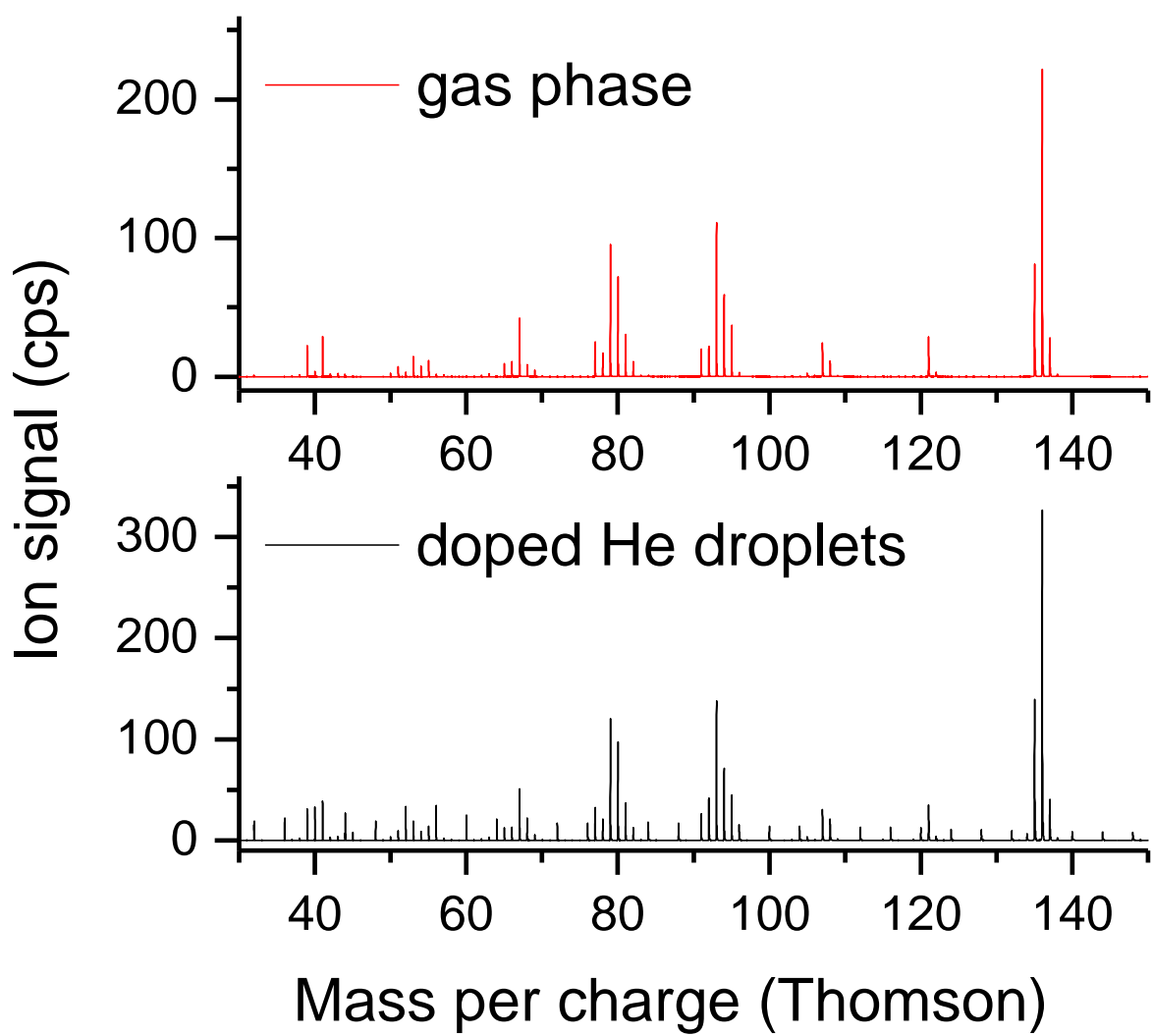

Figure 2: Mass spectra taken in the monomer region of adamantane. Electron energy 80 eV, electron current 29 µA. Top: gas phase adamantane only. Bottom: adamantane doped He droplets, $T_{He}$=9.7 K, $p_{He}$=2.5 MPa. $T_{Adamantane}$ = 310 K, $p_{Adamantane}$ = 1.4 mPa. Note the presence of helium clusters $He_n^+$ in this spectrum.

Prominent features in the gas phase spectrum (top) include the parent cation (m/z 136) and weaker fragment ions with (in order of decreasing intensity) m/z 93, 79, 135, 80, 94, 67, 95 and 41 corresponding to the loss of $C_3H_7$, $C_4H_9$, H, $C_4H_8$, $C_3H_6$, $C_5H_9$, $C_3H_5$ and $C_7H_{11}$, respectively. Apparently, aside from C-H cleavage, the loss of $C_3$ and $C_4$ units is preferred.

The gas-phase spectrum shown in Figure 2 is taken at 80 eV and can be compared with the gas-phase electron impact spectra for adamantane available from the NIST Chemistry WebBook[25] at 70 eV and from SDBS[26] at 75 eV (sample temperature 180°C). There is good agreement with regard to the fragment ions that are observed, but the relative intensities are in better agreement with the SDBS spectrum (our sample temperature is estimated to be 40°C), especially as regards m/z 93.

The spectrum of the doped helium droplets suggests a mode of ionization of adamantane by excited helium species that yields very similar fragments compared to ionization by 80 eV electrons. For example, the excess energy available in the ionization of adamantane (IE = 9.25 +/- 0.04 eV) by $He^+$ (IE = 24.59 eV) is considerable (15.3 eV). The major fragment ions that are observed by electron impact and droplet ionization are identical and their relative intensities are very similar.

No doubly charged adamantane could be observed under the prevailing experimental conditions. The $^{13}C$ isotopologue of adamantane$^{2+}$ was absent. The double ionization threshold for a single molecule of adamantane is $IE_1 + IE_2$. No experimental value for $IE_2$ appears to be available in the literature. Our computations indicate a vertical second ionization energy of around 14.1 eV and an adiabatic value of 11.3 eV, rendering double ionization possible in helium nano-droplets. Geometry optimization suggests that the doubly ionized adamantane deforms without barrier by opening the cage structure. $A_1^{2+}$ molecules are also absent in mass spectra without helium and much higher electron energies. This indicates that they fragment by Coulomb explosion immediately after formation. The large deformation energy associated with the exothermic opening of the cage structure can enhance the fragmentation dynamics. A standard DFT method on the B3LYP/6-311g(d,p) level of theory including Grimme D3 dispersion with Becke Johnson damping was harnessed for all computations with the gaussian 09 software reported in the present work.[27-33] Molecular geometries are given in the supplementary material.

Given the interest expressed in the literature in the inclusion complex He@$C_{10}H_{16}$,[34,35] we should mention that we did not observe the inclusion complex cation He@$C_{10}H_{16}^+$ in our mass spectra. Furthermore, we did not observe He attached to any product ion from adamantane doped He droplets, in contrast to other carbonaceous species embedded in He droplets[36-38] or studies where He was attached to cations in cryogenic traps[39,40] or via co-expansion of ions in seeded beams[20].

**Loss of H Atoms from Adamantane.** Figure 2 shows a significant (A-H)$^+$ peak in both the gas phase and droplet spectra. The H loss channel with He$^+$, reaction (1), is very exothermic, by 14.0 eV, given that IE(He) = 24.6 eV and the appearance energy of the dehydrogenated parent cation AE(A-H)$^+$ = 10.6 eV.[25]

He$^+$ + A → (A-H)$^+$ + H + He  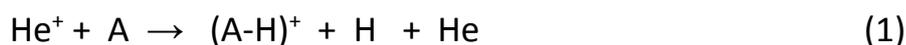  (1)

We see from Figure 3 that H loss persists and even becomes more extensive in the droplet spectra of the adamantane cluster ions. The H loss pattern in the trimer spectrum is representative of the H loss spectra observed for the higher cluster ions.

Figure 3 shows the loss of up to at least 4, perhaps 5, H atoms from the dimer cation and the loss of at least 5 H atoms from the trimer cation, generally with decreasing intensity as the loss increases. The loss of H and 2H (H$_2$?) predominates. The intensity of the dimer cation (A-H)$^+$A at m/z 271 appears anomalously large, larger than the A$_2^+$ dimer cation. (A-H)$^+$A is a cation apparently with special stability, perhaps due to some form of cross bonding. Our DFT calculations starting from intact A and (A-H) moieties led to a stable ((A-H)A)$^+$ structure involving an intermediate H atom with C-H distances of 1.27 Å in a linear C-H-C configuration (Figure S1 in the supplementary material). It has an adiabatic dissociation energy of D$_0$ = 0.65 eV for the reaction (A-H)A$^+$ ⇒ (A-H)$^+$ + A and the positive charge is distributed to A-H (0.34 e) and to A (0.66 e) according to a Mulliken analysis. In another isomer without C-H-C bridge (Figure S2), the charge is located on the (A-H) moiety (0.98 e) and D$_e$ is reduced to 0.24 eV. The cationic parent dimer shows an equally distributed charge on both adamantane moieties and also a rather high binding energy of D$_0$ = 0.63 eV for the reaction $A_2^+ \Rightarrow A^+ + A$. In summary, a distributed charge on both moieties stabilizes the system compared to a localized charge on one of the constituents. A non-negligible contribution to $D_0$ also arises from van der Waals forces; the binding energy of the neutral $A_2$ dimer can be estimated with $D_0$ = 0.14 eV (0.13 eV if corrected for the basis set superposition error using the counterpoise correction).

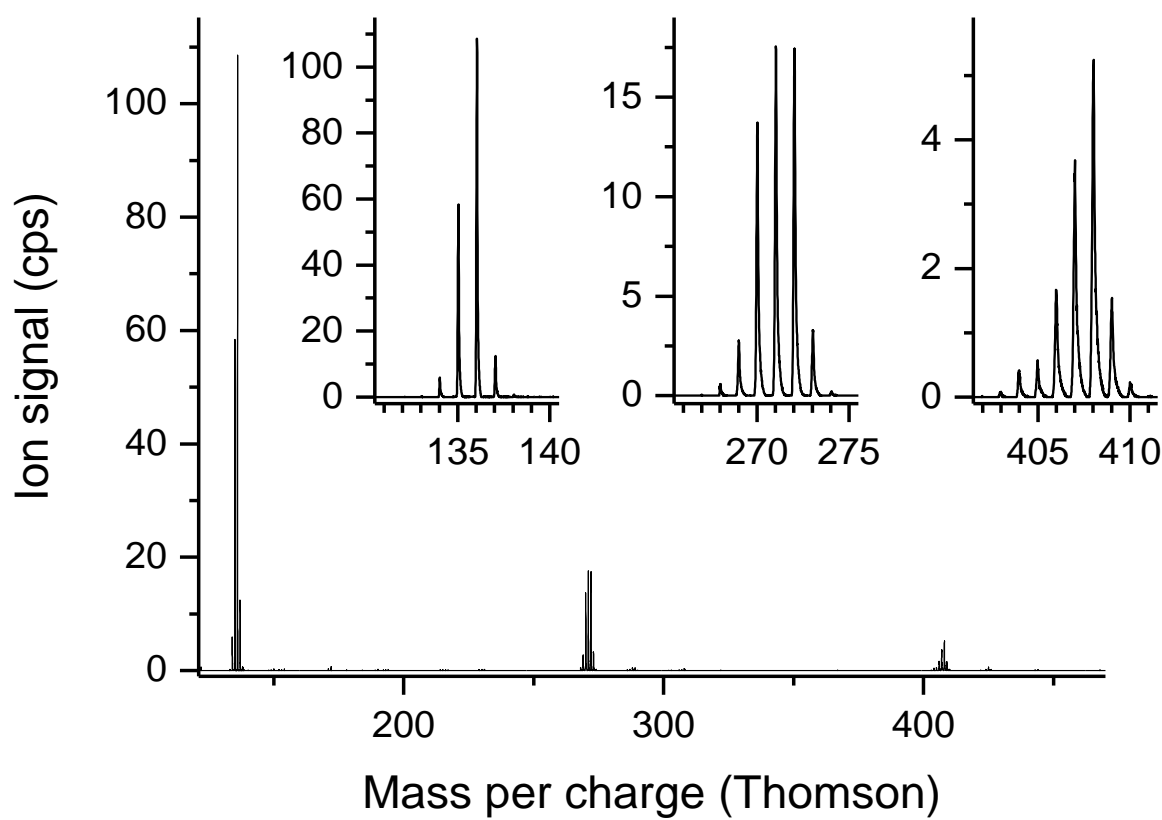

Figure3. Mass spectra taken in the monomer to trimer region of adamantane doped He droplets. Electron energy 80 eV, electron current 29 µA. $T_{He}$=9.7 K, $p_{He}$=2.5 MPa. $T_{Adamantane}$ = 300 K, $p_{Adamantane}$ = 1.4 mPa, Note the signals for helium clusters $He_n^+$ have been subtracted. The inserts show the H loss peaks for the monomer (m/z = 135), dimer (m/z = 271) and trimer (m/z = 407) cations. Further losses of hydrogen are also observed.

**Adamantane Cluster Cations.** The mass spectrum given in Figure 4 shows clearly that helium droplet conditions are well suited for adamantane cluster formation and their mass spectrometric observation after ionization. Huge sizes of singly charged clusters are achieved involving well over 100 (m/z=13600) adamantane molecules, as well as doubly and even triply charged cluster ions. Also we should note the absence, within the limit of detection, of adamantane cluster ions with attached He atoms. Apparently the excess energy deposited in the ionization event prevents the attachment of He atoms under our experimental conditions. In contrast, in the beam expansion experiments mentioned in the introduction in which the He cluster ions were observed, stabilization proceeded by third body collisions.[20]

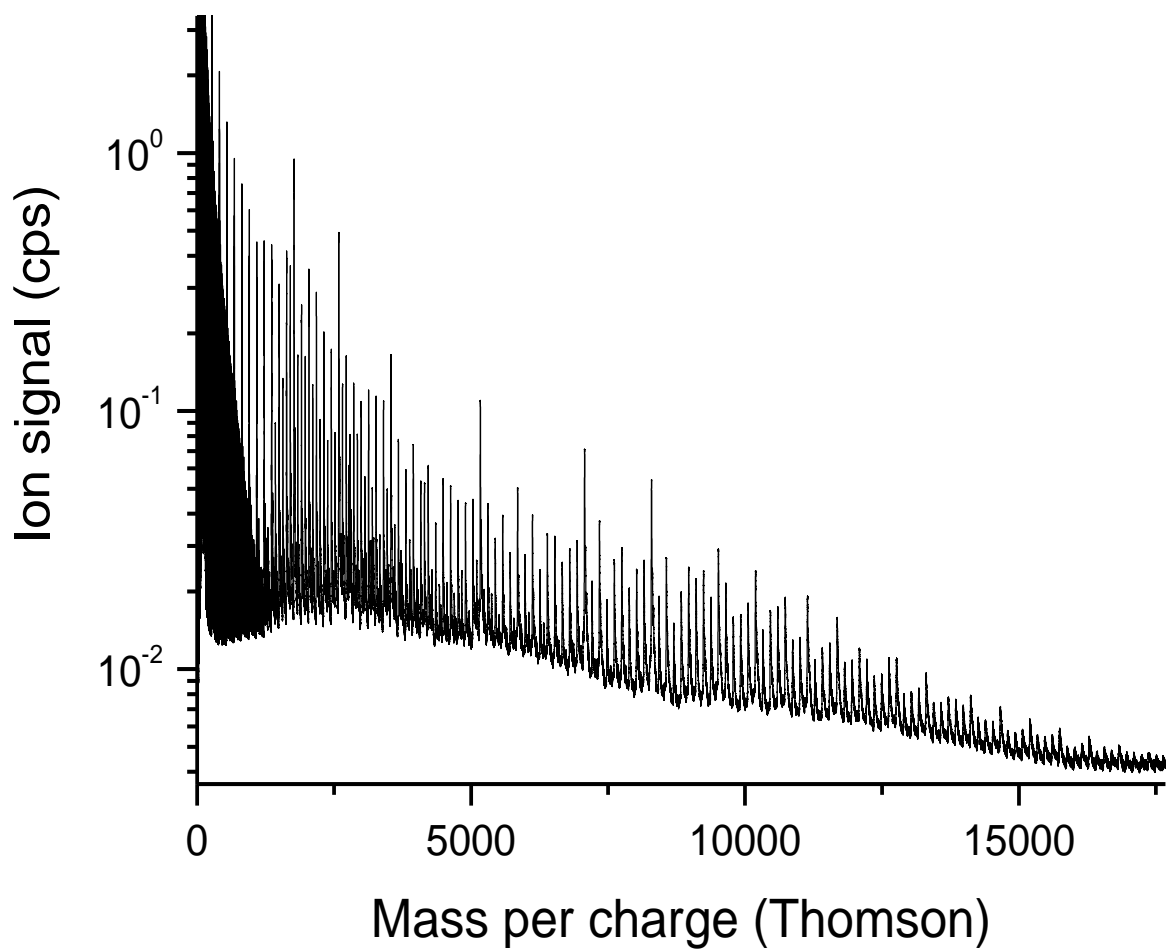

Figure 4. Mass spectrum of helium nanodroplets doped with adamantane. Conditions: $T_{He}$ = 9.7 K, $p_{He}$ = 2.5 MPa, average droplet size $10^6$, $T_{Adamantane}$ = 300K, $p_{Adamantane}$ = 1.9 mPa, $E_{el}$ = 120 eV, $I_{el}$ = 33 µA.

Figure 4 reveals the presence of singly-charged cluster cations of adamantane with itself, $(C_{10}H_{16})_n^+$, with n ranging from 2 to well over 100. Also present are the adamantane clusters of $(A-H)^+A_n$ and $(A-2H)^+A_n$, but in significantly smaller relative amounts. There was no evidence for the formation of the diamantane cation (m/z=190) or higher adamantoid cations.

Pure multiply-charged adamantane clusters could also be resolved, including doubly-charged clusters, $(C_{10}H_{16})_n^{2+}$, with n from 19 to 119 and triply-charged clusters, $(C_{10}H_{16})_n^{3+}$, with n from 52 to 105. The later onsets of these multiply-charged clusters with the size n of the clusters, compared with the onset of singly-charged clusters, can be attributed to Coulomb explosion of the multiply-charged clusters. Apparently Coulomb repulsion between charges renders $(C_{10}H_{16})_n^{2+}$ clusters unstable with n < 19 and $(C_{10}H_{16})_n^{3+}$ clusters already with n < 52.

**Magic Number Adamantane Cluster Cations.** Another striking feature of the adamantane doped nanodroplet mass spectra (see Figure 5) is the presence of anomalous or "magic number" peaks that are more intense than the following peak (by a factor of about 2 or more) for all three charge states. These reflect the presence of cluster ions with special stabilities. The magic numbers are as follows:

1+: 13, 19, 38, 52, 61, 70, 75, 79, 82, 86, 90, 94, 98, 104, 108, 112, 116, 120, 124

2+:         38, 52, 61, 70, 75, 79, 82, 86, 90, 94, 98, 104, 108, 112, 116

3+:                 61, 70, 75, 79, 82, 86, 90, 94, 98

Remarkably, the magic numbers are independent of charge state. Also, some patterns appear to be present: a difference of 4 for n between 75 and 79, 82 and 98 and between 104 and 124, and a difference of 9 between 52 and 79. The anomalously large peaks at n=25, 26 in the dication spectrum in Figure 5 express the beginning of the stability pattern for the doubly charged clusters. In this figure the yield of multiply charged cluster ions that overlap with singly charged cluster ions, i.e. $A_{2n}^{2+}$ or $A_{3n}^{3+}$ and $A_n^+$, were deduced from several mass spectra that were measured at different electron energies. Triply charged cluster ions are exclusively formed at electron energies higher than 40 eV and the yield of doubly charged cluster ions exhibits a much stronger dependence on the electron energy compared to the singly charged ions of similar mass to charge ratio (see Figure 7). Furthermore, the contribution of doubly charged ions can be completely suppressed by changing the pickup conditions to prevent formation of large enough dopant clusters. Figure 6 shows a section of three mass spectra that were taken under conditions that favor (large droplets and high electron energy) or suppress (small droplets and low electron energy) the formation of multiply charged adamantane cluster ions. Please note that the anomalously high yield for the ion designated with $A_{52}^{2+} + A_{26}^+$ in the upper panel of Figure 6 originates exclusively from the contribution of the dication.

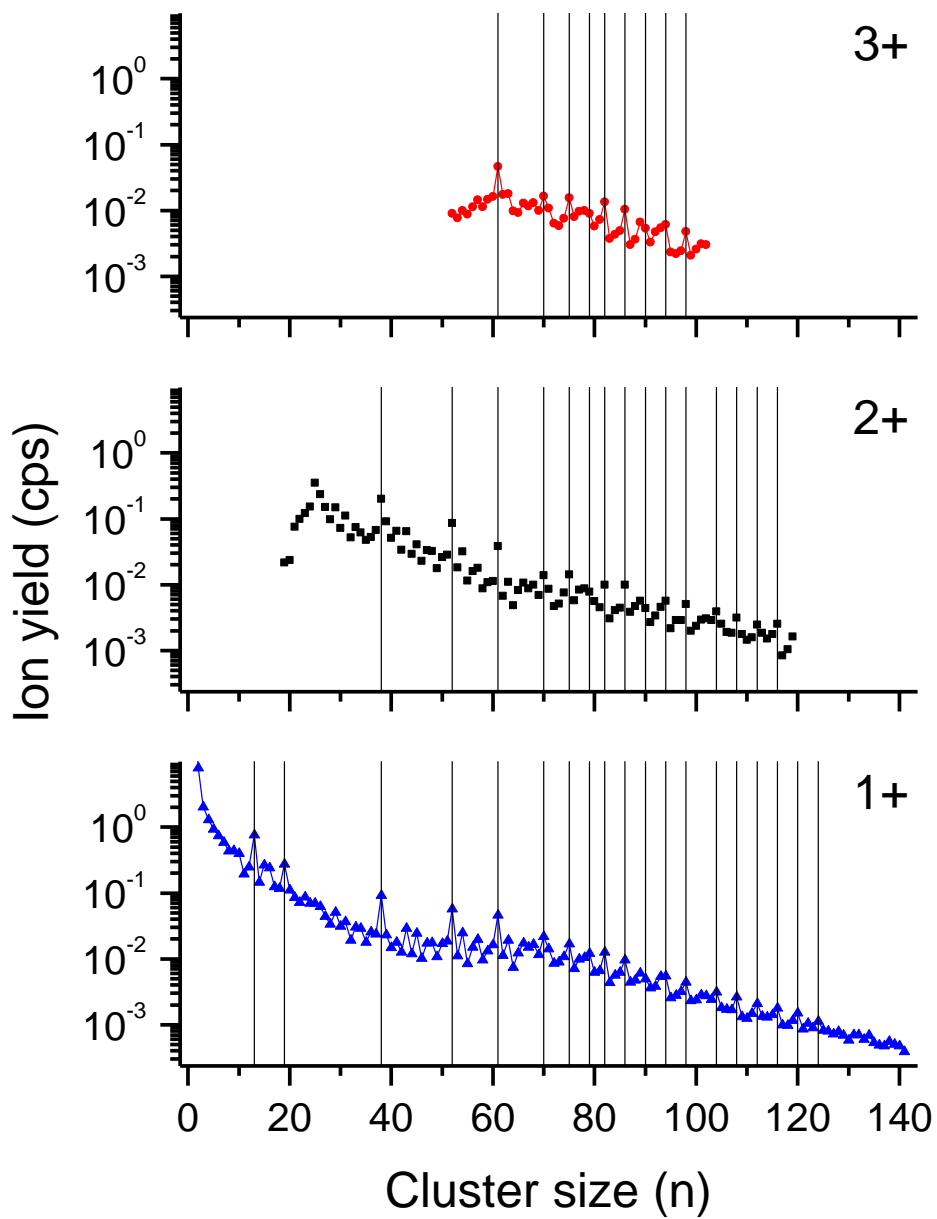

Figure 5. Separated mass spectra for the three charge states of the adamantane clusters in helium nanodroplets doped with adamantane. Conditions: $T_{He}$ = 9.7 K, $p_{He}$ = 2.5 MPa, average droplet size $10^6$, $T_{Adamantane}$ = 300K, $p_{Adamantane}$ = 1.9 mPa, $E_{el}$ = 120 eV, $I_{el}$ = 33 µA.

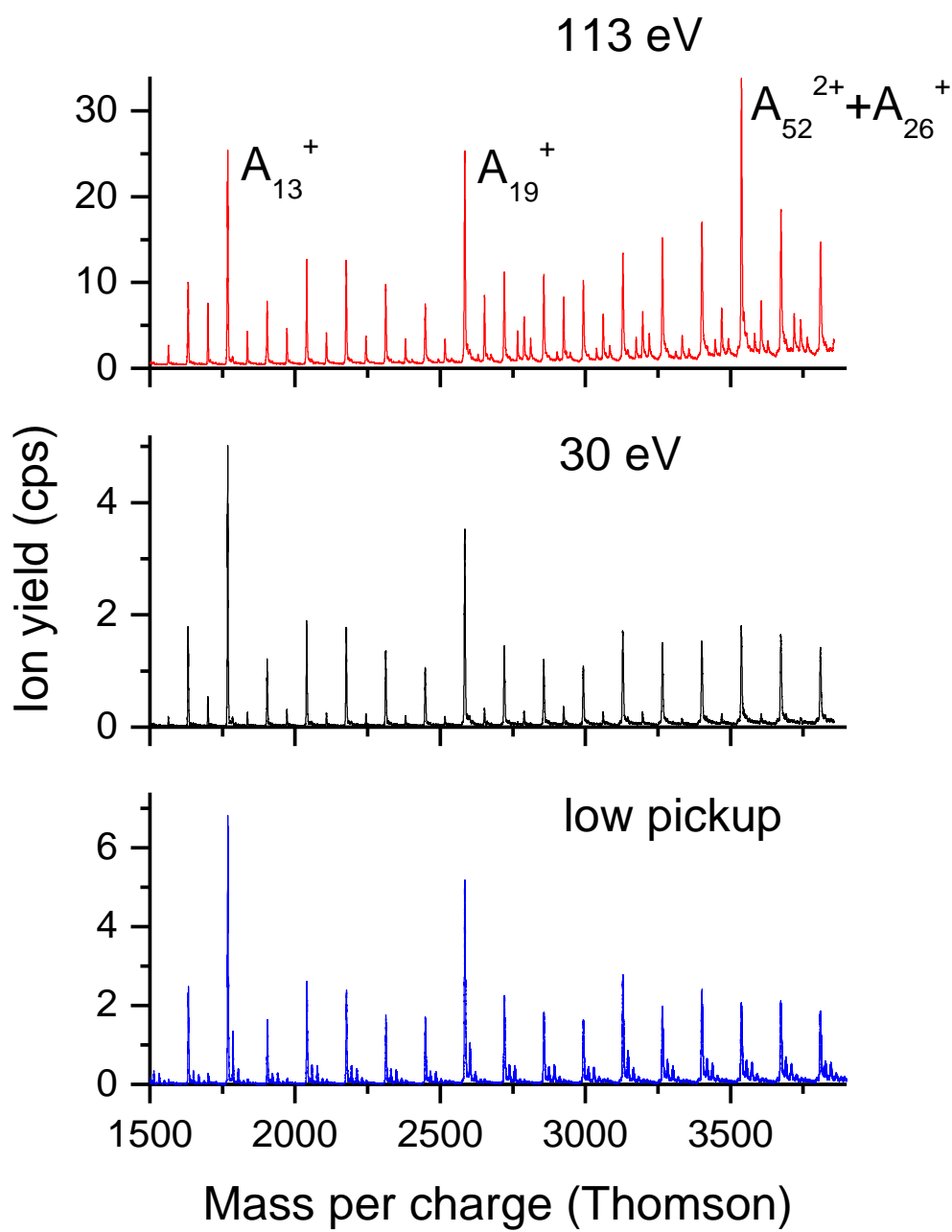

Figure 6. Sections of three mass spectra taken at various conditions to favor or suppress formation of multiply charged cluster ions. Lower panel: (low pickup) $T_{He}$ = 9.9 K, $p_{He}$ = 2.5 MPa, average droplet size $10^5$, $T_{Adamantane}$ = 300K, $p_{Adamantane}$ = 1.9 mPa, $E_{el}$ = 120 eV, $I_{el}$ = 33 μA; center panel (30eV): $T_{He}$ = 9.65 K, $p_{He}$ = 2 MPa, average droplet size $10^6$, $T_{Adamantane}$ = 300K, $p_{Adamantane}$ = 1.9 mPa, $E_{el}$ = 30 eV, $I_{el}$ = 300 μA.; upper panel (113eV): $T_{He}$ = 9.65 K, $p_{He}$ = 2 MPa, average droplet size $10^6$, $T_{Adamantane}$ = 300K, $p_{Adamantane}$ = 1.9 mPa, $E_{el}$ = 113 eV, $I_{el}$ = 300 μA.

Magic numbers of 13 and 19 have been observed previously for singly-charged cluster ions of spherical molecules ($C_{60}$)[41] and atoms (rare gases)[42] for which they have been attributed to packing of a cation that leads to closure of the first icosahedral shell (n = 13) and the formation of a nested icosahedron (n = 19). This seems also to be the case for adamantane. Further packing to form the next icosahedron (n = 55) does not seem to occur with adamantane. Insight into the nature of the structure of the higher magic cluster ions is provided by a very recent computational study of the equilibrium shapes of assemblies of colloidal particles with n between 3 and 85 that interact via an experimentally validated pair potential.[43] Magic number clusters that result from fcc packing were predicted for n = 38, 52, 61, 68, 75 and 79. These correspond exactly with the magic number adamantane clusters observed in our droplet experiments in the three charge states (see Figure 5). The n= 68 signal does not peak as the others do, but there is a sharp drop in signal by almost a factor of 2 in going to n = 69. This perfect agreement of our observed magic numbers for adamantane clusters with shell closures for fcc packing suggests a structural transition of adamantane clusters from icosahedral to fcc packing at a cluster size between n=19 and 38. Diffraction experiments of gas phase adamantane clusters utilizing free electron lasers[44,45] or high resolution transmission electron microscopy measurements of soft-landed clusters[46-48] could provide a direct determination of the structural arrangement of individual clusters.

There is more uncertainty about the nature of the highest cluster cations, especially 82 and 104 that appear to be "core ions" for the +4 sequences from 82 to 94 and 104 to 124. Perhaps 82 and 104 (and some of the others) are a consequence of further FCC packing (the computations in Ref. [43] did not attempt predictions beyond 85) and tetramers of adamantane of special stability (eg. tetrahedrons) prevail in some of the larger adamantane clusters and add to these two core ions.

**Formation of Multiply-Charged Adamantane Cluster Cations.** Multiple ionization can be initiated by the energetic helium species He*(19.8 eV), He$^+$ (24.6 eV) and/or He*$^-$ (19.2 eV) in single or sequential collisions with the neutral molecule or cluster depending on droplet size and the ionizing electron current.[19] The excess energy associated with the ionization event can lead to dissociation and dehydrogenation of the cluster and He evaporization from the droplet. With He$^+$, for example, ionization, dissociative ionization and dehydrogenation can proceed as follows:

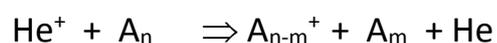

He$^+$ + A$_n$ $\Rightarrow$ A$_{n-m}^+$ + A$_m$ + He

$$\Rightarrow A_{n-m}^{2+} + A_m + He + e$$

$$\Rightarrow (A-H)A_{n-m-1}^{2+} + A_m + H + He + e$$

$$\Rightarrow (A_2-2H)^+ + H_2 + A_{n-2} + e$$

Figure 7 shows the formation of selected mono, di and trications with the variation of the energy of the electrons that are responsible for the production of the energetic helium species He*, He$^+$ and He*$^-$. There is a clear onset for the formation of the singly charged adamantane cluster cation (with n = 22) at IE(He). We also observe a weak signal of singly-charged adamantane clusters at electron energies higher than about 20 eV which indicates that also He* weakly contributes to the formation of these ions. Metastable helium is heliophobic and normally does not interact with heliophilic dopants. However, the heavy doping in the present experiment leads to a massive loss of helium and thus the thin remaining layer of helium can be penetrated by the He*. For dehydrogenated cluster species (except for the dimer) we observe also two threshold energies, however, one at around the ionization energy of He and a second one around 40 eV. With increasing number of hydrogens lost, the contribution with the 40 eV threshold becomes more important. The adamantane cluster dication with (n = 41) has the same onset and so is also produced by He$^+$, but the triply charged adamantane cluster cation (with n = 61) is not. The onset at 40 eV of $A_{61}^{3+}$ and the second onset at 40 eV of $A_{41}^{2+}$ are a consequence of two collisions involving either 2 He* or 2 He*$^-$ or both He* and He*$^-$. There is a hint of a second onset at 45 eV for $A_{61}^{3+}$ which can be attributed to two collisions in which one is He$^+$. Furthermore, the bend towards reduced efficiency for the production of $A_{22}^+$ suggests that single ionization of cluster ions is preempted when He* or He*$^-$ are involved in double or triple ionization. Table S1 in the supplementary information summarizes the threshold energies determined for product ions obtained via electron ionization of helium droplets heavily doped with adamantane.

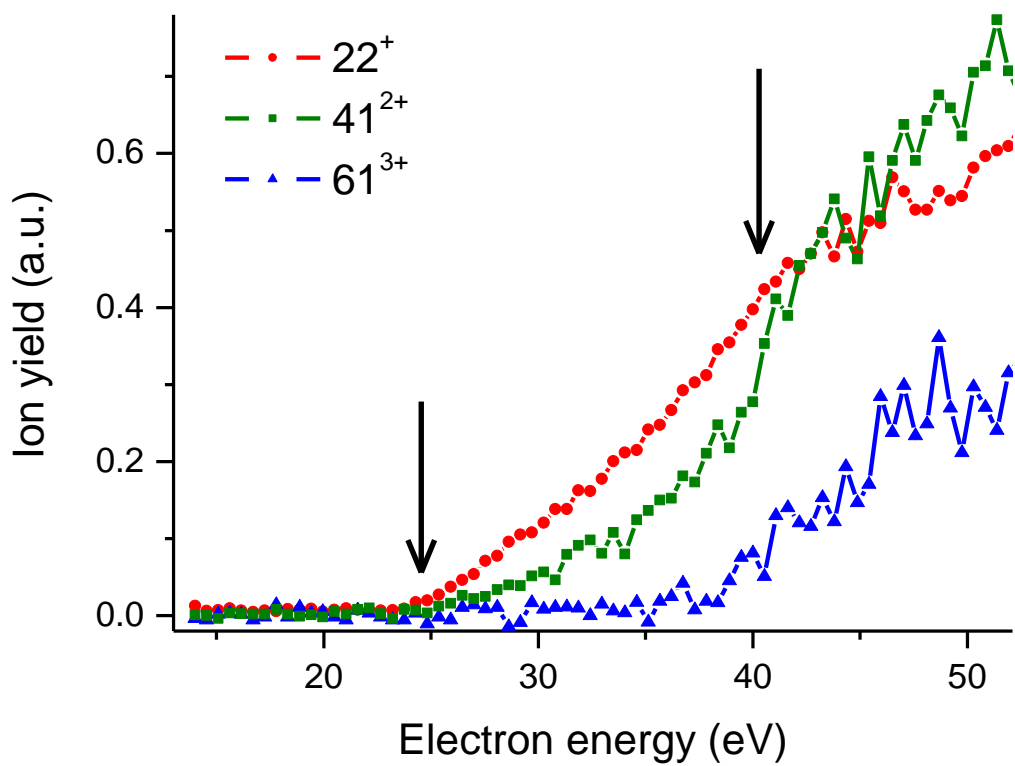

Figure 7. Ion efficiency curves for three selected adamantane cluster ions of different charge states. All curves were normalized at an electron energy of 100 eV. Conditions: helium stagnation pressure: 2 MPa, He temperature: 9.65 K, adamantane pressure 0.16 mPa, electron current 100 µA. The left arrow indicates the ionization energy of He and the right arrow twice the excitation energy of He*.

**CONCLUSIONS**

The aggregation of adamantane, the building block of diamondoids and perhaps diamonds, has been explored in its ionized form near 0 K in He droplets where aggregation is extremely favorable. Aggregation of the first three charge states of adamantane was achieved up to huge cluster sizes (more than 100 molecules) with some dehydrogenation. Single and double collisions with energetic helium species (He*, He$^+$ or He*$^-$) were shown to readily initiate single and multiple ionization of these clusters. The occurrence of Coulomb explosion appeared to discriminate against the formation of small multiply charged clusters. No adamantoid or microdiamond formation was evident in the presence of the charged or excited helium species. Most noteworthy was the tendency of these aggregates to assimilate in structures of special stability that were easily visible as anomalous m/z peaks at "magic number" cluster sizes measured mass spectrometrically. Many of these structures were consistent with previous observations in our laboratory of rare gas clusters, for example, and predictions of stable structures computed by others.

**ACKNOWLEDGEMENTS**

This work was supported by the FWF projects P26635, W1259, M1908. L.K. acknowledges gratefully a grant by the KKKÖ. A.A. and B.R. were supported by an Ernst Mach Follow-up grant and D.K.B. gratefully acknowledges support by York University. The computational results presented have been achieved using the HPC infrastructure LEO of the University of Innsbruck.

**Supporting Information**

Geometries in the format "atom x y z" with length units in Å

Figures of two isomers of dehydrogenated adamantane dimer cations

Table of threshold energies for the product ions formed upon electron ionization of helium droplets heavily doped with adamantane